# Does Supplier Evaluation Impact Process Improvement?

Shiva Prasad H.C.*, Giridhar Kamath, Gopalkrishna Barkur, Rakesh Nayak

*Manipal Institute of Technology, Manipal University (India)*

\* Corresponding author *hcshipra@gmail.com,*

*giridharbk@yahoo.com, gopalkrishna.b@manipal.edu, rakeshnr.16@gmail.com*

**Abstract:**

**Purpose:** The research explores and examines factors for supplier evaluation and its impact on process improvement particularly aiming on a steel pipe manufacturing firm in Gujarat, India.

**Design/Methodology/approach:** The conceptual research framework was developed and hypotheses were stated considering the analysis of literature and discussions with the managers and engineers of a steel pipe manufacturing company in Gujarat, India. Data was collected using in-depth interview. The questionnaire primarily involves the perception of evaluation of supplier. Factors influencing supplier evaluation and its influence on process improvement is also examined in this study. The model testing and validation was done using partial least square method. Outcomes signified that the factors that influence evaluation of the supplier are quality, cost, delivery and supplier relationship management.

**Findings and Originality/value:** The study depicted that quality and cost factors for supplier evaluation are insignificant. The delivery and supplier relationship management have significant influence on evaluation of the supplier. The research also depicted that supplier evaluation has significant influence on process improvement.

**Research limitations/implications:** The study has been made specifically for ABC steel pipe manufacturing industry in Gujarat, India and may not be appropriate to the other industries or any parts of the world. There is a possibility of response bias as the conclusions of this research was interpreted on survey responses taken from the employees of case study company, so it is suggested that future research can overcome this problem by employing various methodologies in addition to surveys like carrying out focus group and in-depth interviews, brainstorming sessions with the experts etc.

***Originality/value:*** Many researchers have considered quality, cost and delivery as the factors for evaluating the suppliers. But for a company it is quintessential to have good relationship with the supplier. Hence, the factor, supplier relationship management is considered for the study. Also, the case study company focused more on quality and cost factors for the supplier evaluation of the firm. However delivery and supplier relationship management are also equally important for a firm in evaluating the supplier.

## 1. Introduction

One of the foremost measures in supply chain is purchasing and procurement (Sarkis & Talluri, 2002). Since the acquisition of goods and services extremely influenced the firm's goal and objectives, organizations are placing high importance on such activities. Successful business firms have developed procurement strategies and policies that are aligned to the company's goals that help the company to uplift their performance in long run (Tahriri, Osman, Ali & Mohd-Yusuff, 2008). A firm plans to meet the demands of the customer in short time and at the lowest manufacturing cost with an aim to be the market leader. Hence, delivery time and total costs reduction are given prominence. Nearly seventy percent of the product's total cost is incorporated by the total cost of the raw materials (Ghodsypour & O'Brien, 1998). By, reducing the cost of raw material the total cost gets reduced, that is influenced by the supplier. Therefore in context with the supply chain, evaluation of the appropriate supplier becomes extremely important.

The research emphasis on factors for the evaluation of the supplier for ABC mild steel pipe manufacturing company in Gujarat, India and its influence on the business process improvement. The major issue for a pipe manufacturing company in India is beginning from the preliminary specification of technical aspects to the placement of the services that is the overall scope of supply chain management. To avoid the grievances from the customer these company have to ensure the timely delivery of the services. Repeated service disrupting results in end user's poor rating that brings down the performance of the business and overall efficiency. For a pipe manufacturing company quality of the product and product life cycle are directly correlated with the evaluation of the supplier. For instance, the purchasing departments need to be accurate in terms of evaluation of the item and supplier evaluation in order to purchase the heavy equipment required for the production. Hence it is very crucial to make sure that the suppliers deliver timely items required as a raw material for the pipe manufacturing and also materials required for maintenance for the heavy equipment (Lai, Yik & Jones, 2006). The supplier must be competent enough and be able to provide sufficient capacity of raw materials continuously during the life span of the supplier. Hence, evaluation of the most appropriate supplier can substantially add value to the performance of the company by effective process of supplier evaluation. To achieve this, the reduction in the cost of product or item with accurate required standard is set so that the there is less frequency in the replacement (Choy & Lee, 2002).

*Research objectives*

1. To determine the factors that influences the supplier evaluation at ABC firm.
2. To develop a model that describes the measure for evaluation of the supplier.
3. To determine the influence of the supplier evaluation on the process improvement.
4. To test and validate the model of supplier evaluation criteria using structural equation model (SEM) technique.

*Research questions*

1. What are the criteria for supplier evaluation at ABC firm?
2. How the decision evaluation of supplier influences the process improvement?
3. Does the correlation among the criteria affect the evaluation of the supplier?

## 2. Literature Review

The process of supplier evaluation becomes a very complicated task as many factors should be taken into account. Dickson (1966) suggested more than 20 factors for the evaluation of the supplier that the purchasing or procurement managers have to consider during the process of evaluation of the supplier (cited in Imeri, 2013). Therefore it is understood that the purchasing managers do a lot more work other than buying goods. The main job of the managers is to make decision considering the important measures along with the persons in the organization. Other than minimizing the cost, the responsibility of the managers of procurement department is to select the appropriate supplier that helps them in accomplishing the wide objectives of the firm.

Interpreting the measures or factors is the first step. Dickson discovered 23 factors for the supplier evaluation. It was based on the questionnaire that was given to 273 American and Canadian procurement managers and agents (Imeri, 2013). The different measures for the evaluation of the supplier discovered by him included quality, price, delivery and relation with the supplier. Ellram and Cooper (1990) also discovered quality, price and rapport with the supplier as the factors for the process of supplier evaluation and applied a hierarchical framework. Additionally in this research all the 23 factors were evaluated through questionnaire and found the significant factors as evolution (Table 1), an approach of partial confirmatory factor analysis to exploratory factors analysis was adopted (further read Fabrigar, Wegener, MacCallum & Strahan, 1999).

| Rank | Factor | Mean Rating | Evolution |
|------|--------|-------------|-----------|
| 1 | Quality | 3.508 | Extreme importance |
| 2 | Delivery | 3.417 | |
| 3 | Warranties and claim policies | 2.998 | |
| 4 | Performance history | 2.849 | |
| 5 | Production facilities and capacity | 2.775 | Considerable importance |
| 6 | Price | 2.758 | |
| 7 | Technical capability | 2.545 | |
| 8 | Financial positions | 2.514 | |
| 9 | Procedural compliance | 2.488 | |
| 10 | Communication system | 2.426 | |
| 11 | Reputation and Position in the industry | 2.412 | |
| 12 | Desire for business | 2.256 | |
| 13 | Management and organization | 2.216 | |
| 14 | Operation controls | 2.211 | Average Importance |
| 15 | Repair services | 2.187 | |
| 16 | Attitude | 2.120 | |
| 17 | Impression | 2.054 | |
| 18 | Packing ability | 2.009 | |
| 19 | Labour relation record | 2.003 | |
| 20 | Geographical location | 1.872 | |
| 21 | Amount of past business | 1.597 | |
| 22 | Training aids | 1.537 | |
| 23 | Reciprocal arrangements | 0.610 | Slight Importance |

Table 1. Dickson factors (Imeri, 2013)

The main aim of the evaluation of the supplier is not only to select the supplier who provides the products for lowest cost but also to see whether the supplier is capable of supplying the products that meets the goal of the firm on continuous basis (Kahraman, Cebeci & Ulukan, 2003). The company to survive in the aggressive market condition has to offer quality products and services as per the demands and needs of the customer. Therefore the organization in order to provide the best product quality and service has to choose the best supplier to make sure they produce a good quality product. Therefore the organization with regard in choosing the appropriate supplier, it's very crucial for an organization to spend time in this prospective. Therefore the decision of the supplier is very significant.

## 3. Research Design

This section mainly focuses on the factors that form the theoretical framework of the study.

### 3.1. Quality Factor

Quality is considered as the most significant factor in the study of the evaluation of the supplier. The most significant construct in the supplier evaluation is perceived as quality of the product or service. Therefore in supplier evaluation process, quality product supplied by the supplier has to be taken in to account carefully as the other influencing factors are also correlated. Product quality of an organization is mainly related to the quality of the raw material supplied by the supplier hence it also meets the requirements of the organization. To gain the confidence of the customer, a firm should meet the needs of the customer by providing the quality of the product that will even exceed the level of expectation of the customer. By taking in to account the above discussions, one of the factors that is included in the process of supplier evaluation is the quality. Thus the above is hypothesized as:

*$H_{01}$: Quality factor of raw material supply has no significant influence on supplier evaluation.*

*$Ha_1$: Quality factor of raw material supply has significant influence on supplier evaluation.*

## 3.2. Cost Factor

Another most important factor in the evaluation of the supplier is cost and price as any firm may enable to manage the operational costs like inventory holding cost, maintenance cost and rework cost if the cost and price factors are employed. 70 percent of the cost of the product is of raw material cost that is the major overhead (Ghodsypour & O'Brien, 1998). To improve the cash flow of the business, a firm should select suitable mechanism for pricing to gain the competitive cost. In the era of wide market competition, it is becoming increasingly critical to choose the best supplier at an ideal price so that the business profits are realized. The cost has been the important critical attribute, the decision of supplier evaluation based only the cost attribute has been approved conventionally by many businesses. Many industries follows the mechanism of calculating the total cost for purchasing a product from different suppliers and the cheapest one is chosen (Ozcan & Suzan, 2011). Thus the factors is hypothesized as:

*$H_{02}$: The cost factor has no significant influence on supplier evaluation.*

*$Ha_2$: The cost factor has significant influence on supplier evaluation.*

## 3.3. Delivery Factor

One of the major attribute in the process of supply chain is timely buying and distribution of the products and services. Late delivery of raw materials causes the delay in production, hence performance of the sales will be poor thus leading to the decreased index in satisfaction of the customer. Punctual delivery is very crucial for a supplier as the supply chain is affected (Vonderembse & Tracey, 1999). Delivery on time is significantly influenced by the lead time. The procurement managers should also consider the performance in the delivery besides the cost factors. Thus the supplier evaluation should be based on delivery rate of the raw material and its influence on the profitability of the company, market shares and satisfaction of the customer (Wilson & Collier, 2000). Thus the hypothesis is stated as:

*$H_{03}$: Responsiveness on the delivery of the supplier has no significant influence on supplier evaluation.*

*$Ha_3$: Responsiveness on the delivery of the supplier has significant influence on supplier evaluation.*

## 3.4. Supplier Relationship Management Factor

For a firm to have a competitive advantage over competitor the significant recognizable strategic attribute is the relationship with the supplier. The strategic position of the firms is enhanced by careful evaluation of the supplier. The attribute in the evaluation of the supplier is to build the strong relationship between the buyer and the seller. It not only improves the business process but also facilitates the communication effectively. Newer technology and high quality products are achieved by effectively managing the relationship with the supplier. The products of the company are enhanced and there is a well-developed effort in research and development if there is commitment and trust among the supplier and buyer in the long term partnership in the business. There is an advantage in the price negotiation due the good relation with the buyer i.e. company that have good relation with the supplier can have benefits such as excellent performance in the delivery, improved service and maintenance and price benefits. Therefore with this effect on the relationship with the supplier it is hypothesized as:

$H_{04}$: *Relationship with the supplier has no significant influence on supplier evaluation.*

$H_{a4}$: *Relationship with the supplier has significant influence on supplier evaluation.*

## 3.5. Process Improvement

A firm's performance is examined by the assessment of the supplier and management of the supply chain (Kannan & Tan, 2002). Partnership with the supplier, customer relationship, quality and sharing of information, time to market is all the measures that evaluate the performance of the firm. To enhance the flexibility of a company, innovative strategies in operations such as lean manufacturing, agile manufacturing, and synchronous manufacturing are introduced. For a firm to be competitive in the long run, improvement in the business processes also plays a crucial role. Evaluation of the supplier often helps a firm to choose the ideal supplier among the various competent suppliers. Quality, cost and punctual delivery and supplier relationship management are the factors on that the supplier evaluation is done that will in turn impact the manufacturing process. Thus the above is hypothesized as:

$H_{05}$: *Supplier evaluation has no significant influence with the Process Improvement.*

$H_{a5}$: *Supplier evaluation has significant influence with the Process Improvement.*

### 3.5.1. Hypothetical Model

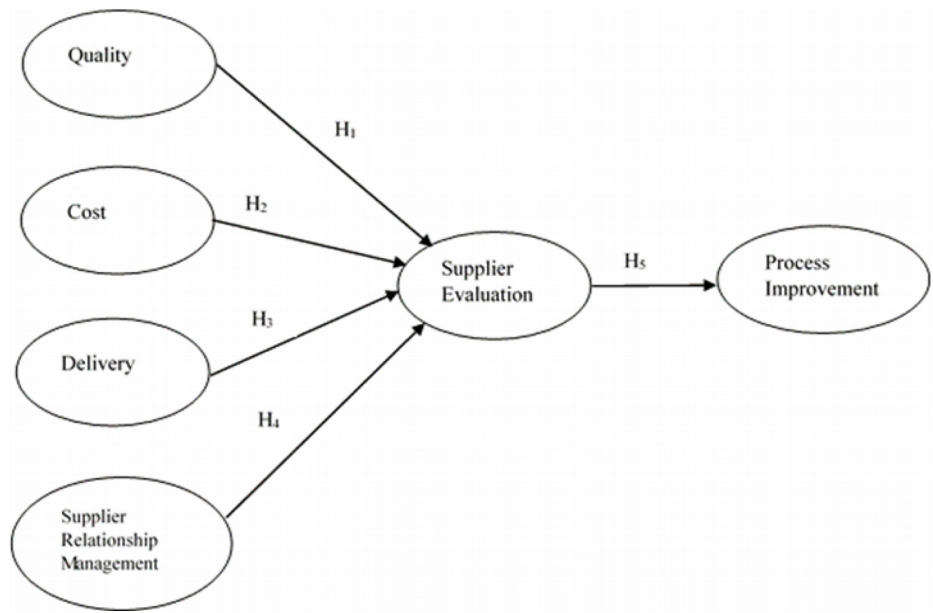

Figure 1. Hypothetical research model

## 4. Methodology

The specific scheme discussed in the paper is for ABC steel pipe manufacturing company in Gujarat, India. The main product of the company is welded Mild steel (MS) pipe. The company also manufactures MS plates and coils, bend pipes and offers coating for the pipes. The main goal of the research is to determine the factors influencing the supplier evaluation and its impact on process improvement in the Indian scenario. The model is evaluated using Partial Least Squares- Structural Equation Modeling (SEM) approach with Smart PLS 2.0. Structural equation modeling is a method used to evaluate a sequence of discrete, interdependent and at the same time also estimate numerous regressions. It is used to obtain relationship between the constructs hypothetically. SEM unlike the other techniques don't have variable limitations. PLS is a tool, used for building models that has numerous factors that are highly collinear. The PLS can handle small sample size hence, this technique was chosen.

The data for the study was collected using the survey technique from the employees of case study company in two stages. The pilot study was carried out to refine the constructs. The research was conducted amongst the small scale of employees of purchasing, logistics, production planning and control, operations departments and the samples is based on convenience. Enhancement of the questionnaire structure by removing the weak items and imperfection in the questionnaire was the main aim of the pilot study. The sample of thirty employees personally participated in the pilot survey and the questionnaire were given personally to all the employees. The original questionnaire comprised of thirty four items. After pilot study was reduced to twenty two items through factor reduction (Table 2). A five points Likert-type scale with one being strongly disagree and five being strongly agree. Thus the information collected from the study was evaluated and analyzed using the Smart Pls 2.0 software. For the final study two hundred samples were considered. The measurement model was tested first and then the structural model so as to evaluate the reliability and validity. The items in the questionnaire is as follows

To verify the latent variables reliability, composite reliability and item reliability measures was calculated. Cronbach's alpha and composite reliability data for the final model is also computed (Table 4). A number of Cronbach's coefficient of alpha values is suggested by various researchers. For the reliability of the data, the minimum threshold value generally taken into account is considered to be 0.6 (Nunnally & Bernstein, 1994). The value of Cronbach's alpha coefficient above 0.9 is excellent (Hair, Black, Babin, Anderson & Tatham, 2006). Any value in between gives the strength of association between variables (Table 3).

| Codes | Constructs |
|---|---|
| QLY1* | Suppliers always provide inspection records for each orders placed. |
| QLY2 | Suppliers fulfil the technical specification that meets requirements of our organization. |
| QLY3 | Products supplied by the supplier are compliant to API and IS specification. |
| QLY4* | Suppliers are certified to the relevant standards. |
| QLY5* | Suppliers provide test certificate for every product for the order placed. |
| QLY6 | Suppliers are committed to continuous improvement of quality. |
| CST1* | Logistics expenses are paid by the supplier. |
| CST2 | Suppliers give discount for bulk orders. |
| CST3 | Faulty products are taken care at the supplier's end. |
| CST4* | Suppliers charge interest, if the payments are made after the credit period. |
| CST5 | Suppliers offer a reasonable credit period to make payments. |
| CST 6* | Suppliers clearly discusses about the future possible price escalations within the contract period. |
| DEL 1 | The delivery mode chosen by the supplier is reliable. |
| DEL 2 | Supplier has a mechanism to track the deliveries. |
| DEL 3 | Supplier promptly replaces the defect product on time. |
| DEL 4* | Suppliers are located close to the FIRM. |
| DEL 5* | Suppliers offer a degree of flexibility in their delivery. |
| DEL 6* | The products that are received in the plant are in good condition. |
| SRM 1* | Supplier allow our firm to conduct audit in their workplace. |
| SRM 2 | The suggestions given by the FIRM are readily implemented by the suppliers. |
| SRM 3* | The background reputation of the supplier is taken into account before placing the order. |
| SRM 4 | Suppliers are willing to share sensitive information. |
| SRM 5 | Suppliers are selected on the basis of their long term relationship with the organization. |
| SRM 6* | Suppliers involve buying companies representatives in their Research and Development activities. |
| SEN 1 | Suppliers deliver high quality products. |
| SEN 2 | Suppliers offer reasonable product delivery time. |
| SEN 3 | Suppliers offer products at a competitive prices. |
| SEN 4 | Suppliers build a level of trust and interact effectively with the firm. |
| PIM 1 | Supplier evaluation helps the firm to simplify its process by eliminating wasteful re-doings. |
| PIM 2 | Supplier evaluation helps our company in prevention of the defect. |
| PIM 3 | Supplier evaluation helps the firm in improving the standards. |
| PIM 4* | Supplier evaluation helps in eliminating wastes. |
| PIM 5* | Supplier evaluation helps the firm to eliminate root cause problem relevant to productivity. |
| PIM 6* | Supplier evaluation helps in reducing the process set up time. |

* Indicates the constructs that are eliminated after pilot study.

Table 2. Questionnaire

| '∝' coefficient range | Strength of association |
|---|---|
| <0.6 | Poor |
| 0.6 to 0.7 | Moderate |
| 0.7 to 0.8 | Good |
| 0.8 to 0.9 | Very Good |
| 0.9 | Excellent |

Table 3. Cronbach alpha coefficient (Hair, 2006)

The outcomes of the convergent validity is evaluated on the results of composite reliability and factor loadings that is indicated by moderate to high acceptable range of factor loading for all items and is considered good composite reliabilities. The factor loading values are considered to be greater than 0.4 (Wong, 2013). The least factor loadings for the model being 0.551 and the highest being 0.921 (Table 6). To test the model for discriminant validity, the square root of average variance extracted (AVE) is compared with the correlation between the construct and the other constructs. It is observed that the square root of average variance extracted is greater when compared to the correlation that exists between the constructs (Table 5).

## 4.1. Checking for Reliability and Validity

*For Reliability Testing:*

- Internal consistency Reliability - Cronbach's Alpha values > = 0.6 (Hair, 2006)

*Validity Testing:*

1. Convergent Validity - AVE values must be > = 0.5 (Wong, 2013; Bagozzi & Yi, 1988)
2. Discriminant Validity - square root AVE value must be greater than other (Wong, 2013; Fornell & Larcker, 1981).

|  | AVE | Composite Reliability | R Square | Cronbach Alpha | Communality | Redundancy |
|---|---|---|---|---|---|---|
| QLY | 0.622 | 0.831 | 0 | 0.696 | 0.622 | 0 |
| CST | 0.668 | 0.857 | 0 | 0.751 | 0.668 | 0 |
| DEL | 0.606 | 0.820 | 0 | 0.714 | 0.606 | 0 |
| SRM | 0.502 | 0.746 | 0 | 0.604 | 0.502 | 0 |
| SEN | 0.578 | 0.846 | 0.624 | 0.759 | 0.578 | 0.050 |
| PIM | 0.738 | 0.893 | 0.379 | 0.822 | 0.738 | 0.275 |

Table 4. Reliability measures

|  | QLY | CST | DEL | SRM | SEN | PIM |
|---|---|---|---|---|---|---|
| QLY | **0.788** | 0 | 0 | 0 | 0 | 0 |
| CST | 0.412 | **0.818** | 0 | 0 | 0 | 0 |
| DEL | 0.439 | 0.248 | **0.779** | 0 | 0 | 0 |
| SRM | 0.340 | 0.368 | 0.202 | **0.709** | 0 | 0 |
| SEN | 0.474 | 0.424 | 0.607 | 0.599 | **0.761** | 0 |
| PIM | 0.335 | 0.390 | 0.333 | 0.644 | 0.616 | **0.859** |

Table 5. Latent Variable Correlations and discriminant validity

|  | QLY | CST | DEL | SRM | SEN | PIM |
|---|---|---|---|---|---|---|
| QLY 2 | 0.778 | 0 | 0 | 0 | 0 | 0 |
| QLY 3 | 0.792 | 0 | 0 | 0 | 0 | 0 |
| QLY 6 | 0.796 | 0 | 0 | 0 | 0 | 0 |
| CST 2 | 0 | 0.791 | 0 | 0 | 0 | 0 |
| CST 3 | 0 | 0.746 | 0 | 0 | 0 | 0 |
| CST 5 | 0 | 0.907 | 0 | 0 | 0 | 0 |
| DEL 1 | 0 | 0 | 0.881 | 0 | 0 | 0 |
| DEL 2 | 0 | 0 | 0.681 | 0 | 0 | 0 |
| DEL 3 | 0 | 0 | 0.761 | 0 | 0 | 0 |
| SRM 2 | 0 | 0 | 0 | 0.551 | 0 | 0 |
| SRM 4 | 0 | 0 | 0 | 0.853 | 0 | 0 |
| SRM 5 | 0 | 0 | 0 | 0.690 | 0 | 0 |
| SEN 1 | 0 | 0 | 0 | 0 | 0.747 | 0 |
| SEN 2 | 0 | 0 | 0 | 0 | 0.713 | 0 |
| SEN 3 | 0 | 0 | 0 | 0 | 0.791 | 0 |
| SEN 4 | 0 | 0 | 0 | 0 | 0.788 | 0 |
| PIM 1 | 0 | 0 | 0 | 0 | 0 | 0.739 |
| PIM 2 | 0 | 0 | 0 | 0 | 0 | 0.921 |
| PIM 3 | 0 | 0 | 0 | 0 | 0 | 0.905 |

Table 6. Factor Loadings after Reduction

## 4.2. Descriptive Analysis

### 4.2.1. Quality (QLY)

Here the responses of the employees to the quality are discussed (Table 7). The strongest response was observed for the variable: 'Suppliers fulfil the technical specification that meets requirements' with mean and standard deviation (4.455, 0.632) and weakest response for: 'Suppliers are committed to continuous improvement of quality' with mean and standard deviation (3.83, 0.182). Majority of the employees have responded as good (43.8%) followed by very good (33%) and 20.5% has considered this dimension as average. 2.7 % have responded to this factors as bad and poor being 0% (Figure 2).

|  | Mean | Standard deviation | Poor (1) % | Bad (2) % | Avg (3) % | Good (4) % | V Good (5) % |
|---|---|---|---|---|---|---|---|
| QLY 2 | 4.455 | 0.632 | 0.0 | 0.0 | 7.5 | 39.5 | 53.0 |
| QLY 3 | 3.930 | 0.793 | 0.0 | 4.0 | 23.0 | 49.0 | 24.0 |
| QLY 6 | 3.830 | 0.815 | 0.0 | 4.0 | 31.0 | 43.0 | 22.0 |
| **Average** | **4.072** | **0.747** | **0.0** | **2.7** | **20.5** | **43.8** | **33.0** |

Table 7. Quality factor responses

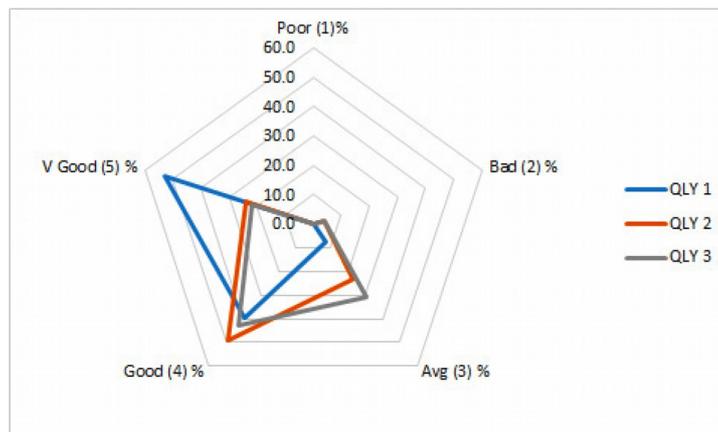

Figure 2. Radar diagram on quality responses

## 4.2.2. Cost

The responses of the employees to the cost factors are discussed (Table 8). The strongest response was observed for the variable: 'Suppliers offer a reasonable credit period to make payments 'with mean and standard deviation (4.105, 0.905) and weakest response for: 'Faulty products are taken care at the supplier's end' with mean and standard deviation (3.875, 1.138). Majority of the employees have responded as good (38%) followed by very good (37%) and 15.5% has considered this dimension as average. 8.8% have responded to this factors as bad and poor being 0.7% (Figure 3).

|  | Mean | Standard deviation | Poor (1) % | Bad (2) % | Avg (3) % | Good (4) % | V Good (5) % |
|---|---|---|---|---|---|---|---|
| CST 2 | 4.075 | 0.826 | 0.0 | 2.0 | 24.5 | 37.5 | 36.0 |
| CST 5 | 4.105 | 0.905 | 0.0 | 7.5 | 13.5 | 40.0 | 39.0 |
| CST 3 | 3.875 | 1.138 | 2.0 | 17.0 | 8.5 | 36.5 | 36.0 |
| **Average** | **4.018** | **0.956** | **0.7** | **8.8** | **15.5** | **38.0** | **37.0** |

Table 8. Cost factor responses

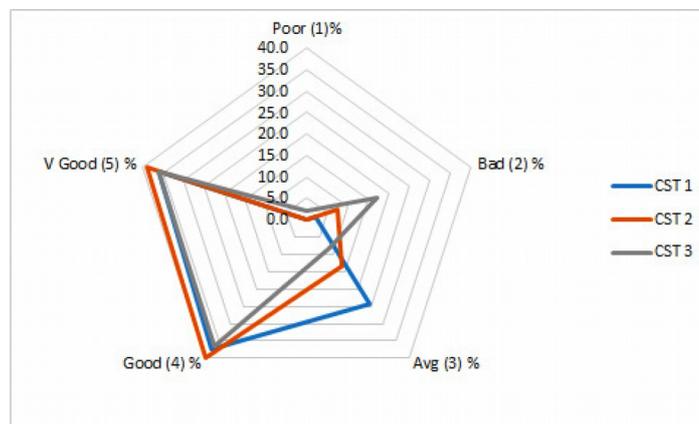

Figure 3. Radar diagram on cost responses

## 4.2.3. Delivery

Here the responses of the employees to the delivery factor are considered (Table 9). The strongest response was observed for the variable: 'The delivery mode chosen by the supplier is reliable' with mean and standard deviation (3.99, 0.8012) and weakest response for: 'Supplier promptly replaces the defect product on time' with mean and standard deviation (3.69, 1.140). Majority of the employees have responded as good (40.5%) followed by very good (29%) and 20.5% has considered this dimension as average. 9 % have responded to this factors as bad and poor being 1%. (Figure 4).

|  | Mean | Standard deviation | Poor (1) % | Bad (2) % | Avg (3) % | Good (4) % | V Good (5) % |
|---|---|---|---|---|---|---|---|
| DEL 2 | 3.945 | 0.898 | 0.0 | 8.5 | 17.5 | 45.0 | 29.0 |
| DEL 3 | 3.690 | 1.140 | 3.0 | 15.5 | 20.5 | 31.5 | 29.5 |
| DEL 1. | 3.990 | 0.802 | 0.0 | 3.0 | 23.5 | 45.0 | 28.5 |
| **Average** | **3.875** | **0.947** | **1.0** | **9.0** | **20.5** | **40.5** | **29.0** |

Table 9. Delivery factor responses

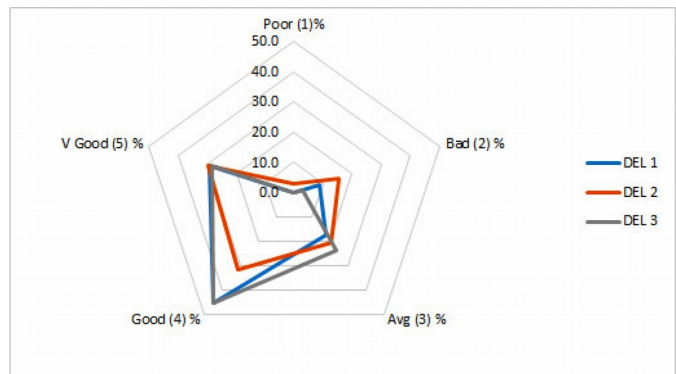

Figure 4. Radar diagram on delivery responses

### 4.2.4. Supplier Relationship Management (SRM)

The response of the employees to the SRM factor is depicted (Table 10). The strongest response was observed for the variable: 'The background reputation of the supplier is taken into account before placing the order 'with mean and standard deviation (4.105, 0.905) and weakest response for: 'Suppliers are selected on the basis of their long term relationship with the firm' with mean and standard deviation (3.875, 1.138). Majority of the employees have responded as very good (33.83%) followed by good (33.67%) and 16% has considered this dimension as average. 10.53% have responded to this factors as bad and poor being 5.67% (Figure 5).

|  | Mean | Standard deviation | Poor (1) % | Bad (2) % | Avg (3) % | Good (4) % | V Good (5) % |
|---|---|---|---|---|---|---|---|
| SRM 4 | 3.595 | 1.161 | 1.50 | 23.00 | 17.50 | 30.50 | 27.50 |
| SRM 3 | 4.330 | 0.897 | 3.00 | 2.00 | 5.00 | 39.00 | 51.00 |
| SRM 5 | 3.450 | 1.271 | 12.50 | 7.50 | 25.50 | 31.50 | 23.00 |
| **Average** | **3.792** | **1.120** | **5.67** | **10.83** | **16.00** | **33.67** | **33.83** |

Table 10. Supplier Relationship Management factor responses

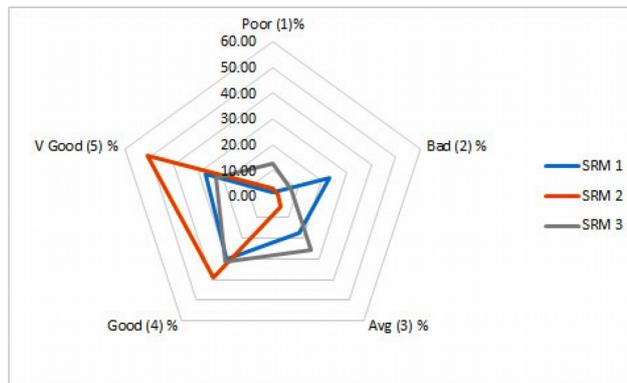

Figure 5. Radar diagram on SRM responses

### 4.2.5. Supplier Evaluation

The response of the employees to the Supplier evaluation factor is depicted (Table 11). The strongest response was observed for the variable: 'Suppliers build a level of trust and interact effectively with the firm 'with mean and standard deviation (4.34, 0.829) and weakest response for: 'Suppliers offer products at a competitive prices' with mean and standard deviation (3.975, 1.138). Majority of the employees have responded as good (41.67%) followed by very good (33.17%) and 16.50% has considered this dimension as average. 4.67% have responded to this factors as bad and poor being 4.00% (Figure 6).

|  | Mean | Standard deviation | Poor (1) % | Bad (2) % | Avg (3) % | Good (4) % | V Good (5) % |
|---|---|---|---|---|---|---|---|
| SEN 1 | 3.935 | 0.978 | 3.00 | 6.00 | 15.00 | 46.50 | 29.50 |
| SEN 2 | 3.740 | 1.135 | 6.00 | 8.00 | 20.50 | 37.00 | 28.50 |
| SEN 3 | 4.185 | 0.892 | 3.00 | 0.00 | 14.00 | 41.50 | 41.50 |
| SEN 4 | 4.340 | 0.829 | 1.50 | 0.00 | 14.00 | 32.00 | 52.50 |
| **Average** | **3.953** | **1.001** | **4.00** | **4.67** | **16.50** | **41.67** | **33.17** |

Table 11. Supplier evaluation factor responses

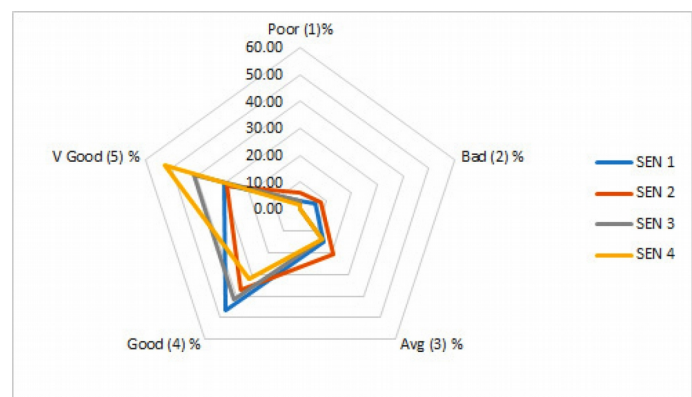

Figure 6. Radar diagram on supplier evaluation responses

## 4.2.6. Process Improvement

The response of the employees to the Process improvement is discussed (Table 12). The strongest response was observed for the variable: 'Supplier evaluation helps the firm in improving the standards' with mean and standard deviation (4.105, 0.905) and weakest response for: 'Supplier evaluation helps the firm to simplify its process by eliminating wasteful re-doings' with mean and standard deviation (3.875, 1.138). Majority of the employees have responded as very good (50.5%) followed by good (34%) and 12.2% has considered this dimension as average. 1.3% have responded to this factors as bad and poor being 2% (Figure 7).

|  | Mean | Standard deviation | Poor (1) % | Bad (2) % | Avg (3) % | Good (4) % | V Good (5) % |
|---|---|---|---|---|---|---|---|
| PIM 2 | 4.445 | 0.819 | 1.5 | 0 | 12 | 25.5 | 61 |
| PIM 3 | 4.355 | 0.826 | 1.5 | 2 | 7.5 | 37.5 | 51.5 |
| PIM 1 | 4.090 | 0.952 | 3 | 2 | 17 | 39 | 39 |
| **Average** | **4.297** | **0.867** | **2.0** | **1.3** | **12.2** | **34.0** | **50.5** |

Table 12. Process improvement factor responses

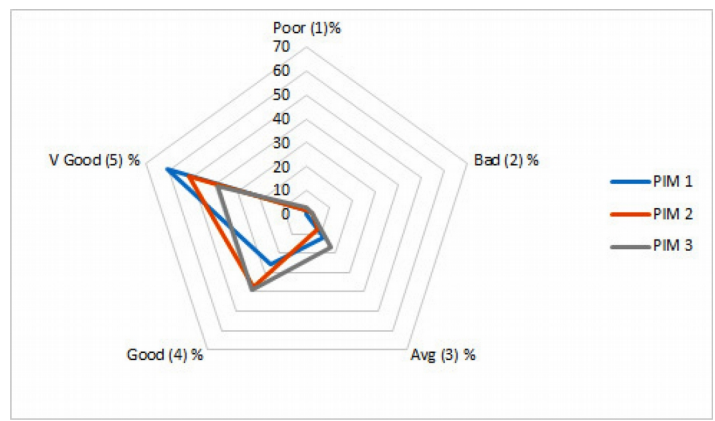

Figure 7. Radar diagram on process improvement responses

## 5. Results

### 5.1. The Structural Model

The model consists of five variables to be tested for their relationships that are designed to test five hypothesis (Figure 1). Path coefficients indicates the relationship strength between the two latent variables. The path coefficient and the exploratory power ($R^2$) are computed for the hypothesized model (Figure 8).

Explanation of variance in target endogenous constructs.

Process improvement (PIM) is an endogenous latent variable. Its co-efficient of determination $R^2$ is 0.379. This means one latent variable, supplier evaluation (SEN) explains 37.9% of the variance in process improvement (PIM). The quality (QLY), cost (CST), delivery (DEL) and relation with the supplier (SRM) together explain 62.4% of the variance of Supplier evaluation (SEN). Path Coefficient sizes of inner model and its significance.

- The inner model suggests that delivery (DEL) has the strongest effect on supplier evaluation (0.455), followed by supplier relationship management (0.437), cost (0.119) and quality (0.077).

- The delivery has the strongest effect on supplier evaluation (0.613).

- The hypothesized path relationship between quality and supplier evaluation is not statistically significant as path coefficient (0.077) is less than 0.1 (Wong, 2013).

- The hypothesized path relationship between cost and supplier evaluation is statistically significant (path coefficient 0.119).

- The hypothesized path relationship between delivery and supplier evaluation is statistically significant (path coefficient 0.455).

- The hypothesized path relationship between supplier relationship management and Supplier evaluation is statistically significant (path coefficient 0.437).

- The hypothesized path relationship between supplier evaluation and process improvement is statistically significant (path coefficient 0.616).

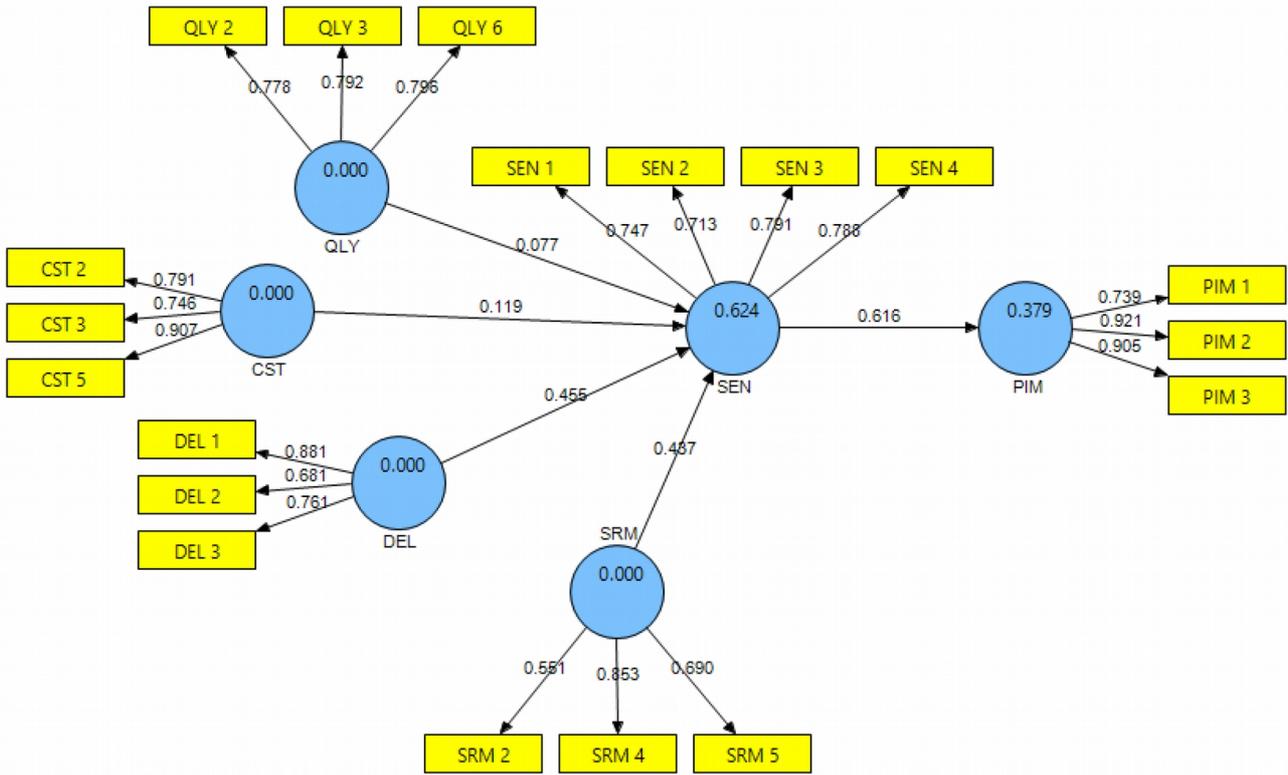

Figure 8. Path coefficient

## 5.2. The Measurement Model

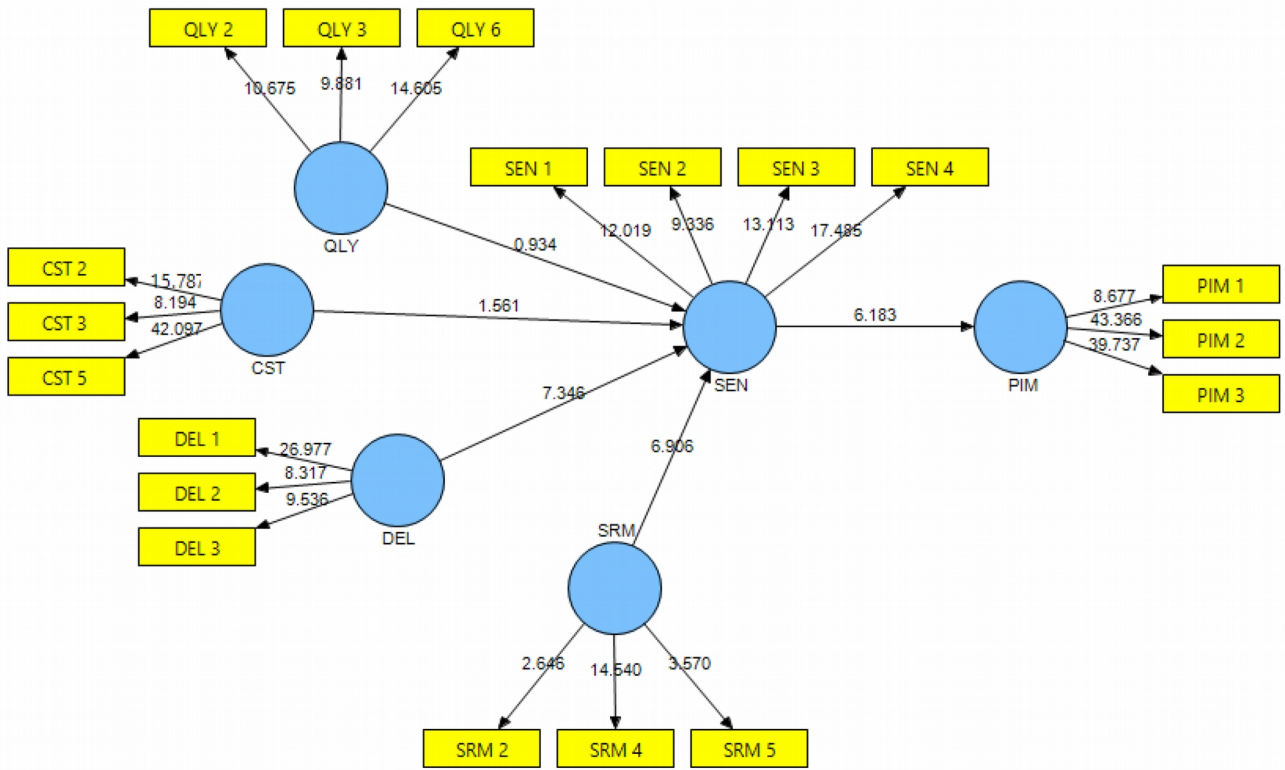

Figure 9. Operational measurement model

| | Original Sample (O) | Sample Mean(M) | Std.Dev. (STDEV) | Std.Error (STERR) | T Statistics (|O/STERR|) | Hypothesis Supported |
|---|---|---|---|---|---|---|
| QLY → SEN | 0.077 | 0.081 | 0.083 | 0.083 | 0.934 | $H_{01}$ |
| CST → SEN | 0.119 | 0.132 | 0.076 | 0.076 | 1.561 | $H_{02}$ |
| DEL → SEN | 0.455 | 0.443 | 0.062 | 0.062 | 7.346 | $H_{a3}$ |
| SRM → SEN | 0.437 | 0.438 | 0.063 | 0.063 | 6.906 | $H_{a4}$ |
| SEN → PIM | 0.616 | 0.607 | 0.100 | 0.100 | 6.183 | $H_{a5}$ |

Table 13. Hypothesis relationship

For hypothetical model being tested, three hypothesis are supported (Table 13). The t-values indicates the significance of relationships that enable the testing of hypothesis. The hypotheses is supported when the t-values is above 1.96 for 0.05 level of significance (Gefen, Straub & Boudreau, 2000).

The results depicted that the two of five hypothesis are not supported. The remaining three were found to be significant statistically.

Following alternate hypotheses are accepted:

*$Ha_3$: Performance and responsiveness on the delivery of the supplier has significant influence on supplier evaluation at 7.436.*

*$Ha_4$: Relationship with the supplier has significant influence on supplier evaluation at 6.906.*

*$Ha_5$: Supplier evaluation has significant influence with the Process Improvement at 6.183.*

*The following null hypothesis are accepted:*

*$H_{01}$: Quality factor of raw material supply has no significant influence on supplier evaluation.*

*$H_{02}$: The cost factor has no significant influence on supplier evaluation.*

## 6. Discussions

With respect to the study conducted in ABC steel pipe manufacturing industry in Gujarat, India; delivery (DEL) and supplier relationship management (SRM) have significant influence on supplier evaluation (SEN). Whilst the quality (QLY) and cost (CST) don't have significant relationships on supplier evaluation (SEN). The result also indicated that the supplier evaluation have significant influence on process improvement (PIM).

The result depicted that there is no significant relationship between quality and supplier evaluation. This depicts that if a supplier is providing high quality products there might be less chances of supplier taken into consideration. Many firms demand the right quality level for their product, therefore if a supplier provides the product with the quality that exceeds the requirement of the firm then there is chances that the supplier being not selected. Another reason for this is the firm have to pay more for the high quality product that in turn results the end customer paying more for the firm's end product. The study depicted that the cost becomes insignificant relationship with the supplier evaluation. Irrespective of the supply chain position of a firm, the cost is given the least importance in the process of supplier evaluation (Choi & Hartley, 1996). Cost has no significant relationship as the firm will naturally opt for only those suppliers, supplying the raw material at lowest cost. Thus if a supplier is providing raw material for the lower cost there is less chances of evaluating the suppliers.

Delivery has significant relationship on supplier evaluation. The strategic goal of steel pipe industry is timely delivery of the product. Therefore the firm to be successful in the competitive market, timely delivery of the product is crucial. The firm can solve problems related to delay in delivery of the product if the suppliers provide the raw materials on time. This findings is also supported in the study narrated by Choi and Hartley (1996). The study also resulted that supplier relationship management has significant influence on supplier evaluation. By maintaining the good relationship with the supplier the firm may stimulate the trust with the supplier and may also promote the two way communication and sharing of information among the supplier and the company. The company can also involve the supplier in their research and development by establishing the long term relationship with the firm. This findings is also supported in the study narrated about the relationship of seller-buyer according to Hunt and Morgan (1995); Lyons & Mehta (1997). The study depicted that the supplier evaluation has positive influence on process improvement. By choosing the best suppliers who provides good quality raw materials, there is improvement in the process of manufacturing like root cause problem elimination, defect prevention and improving standards in the production. This helps the company to expand the production capacities and thereby enhancing the company's profits. This finding is supported by the authors Currie, Dessai, Khan, Wang and Weerakkody (2003).

## 7. Implications and Conclusions

The relationships between the variables are positive or negative. There can be either positive or negative relationship between the variables. The above model has proven this concept. The primary goal of the research is to depict the relationship amongst the dimensions quality, delivery, cost and supplier relationship management on supplier evaluation and supplier evaluation on process improvement. Delivery and supplier relationship management have positive and significant influence on Supplier evaluation. Quality and cost have weak and insignificant impact on supplier evaluation. Supplier evaluation has a positive and significant influence on process improvement. For a firm to have an edge over its competitors as well as to retain its customers it is important to identify the variables that can cause a negative impact and initiate actions so as to neutralize its effect. Failing to do so can result in downfall and there is a possibility of customers migrating to the competitors. Customers expect more, have more choices and are lesser brand loyal these days that is due to the result of competitors offering similar products at a lesser price. To survive in the market the organization has to continually improve its products and services as well meet the unmet needs of the customer.